\documentclass[9pt,conference]{IEEEtran}
\IEEEoverridecommandlockouts

\usepackage{cite}
\usepackage{amsmath,amssymb,amsfonts}
\usepackage{algorithmic}
\usepackage{graphicx}
\usepackage{textcomp}
\usepackage{xcolor}
\usepackage{multirow}
\usepackage{booktabs}
\usepackage{arydshln}
\usepackage{hyperref}
\usepackage{comment}
\begin{document}

\title{
Neural Ambisonic Encoding For Multi-Speaker Scenarios Using A Circular Microphone Array\\
\thanks{*This work was conducted during Yue Qiao's internship at Tencent AI Lab, Bellevue, USA.}
}

\author{\IEEEauthorblockN{Yue Qiao}
\IEEEauthorblockA{\textit{3D3A Lab, Princeton University} \\
Princeton, USA \\
yqiao@princeton.edu}
\and
\IEEEauthorblockN{Vinay Kothapally}
\IEEEauthorblockA{\textit{Tencent AI Lab} \\
Bellevue, USA \\
{vkothapally@global.tencent.com}}
\and
\IEEEauthorblockN{Meng Yu}
\IEEEauthorblockA{\textit{Tencent AI Lab} \\
Bellevue, USA \\
{raymondmyu@global.tencent.com}}
\and
\IEEEauthorblockN{Dong Yu}
\IEEEauthorblockA{\textit{Tencent AI Lab} \\
Bellevue, USA \\
{dyu@global.tencent.com}}
}

\maketitle

\begin{abstract}

Spatial audio formats like Ambisonics are playback device layout-agnostic and well-suited for applications such as teleconferencing and virtual reality. Conventional Ambisonic encoding methods often rely on spherical microphone arrays for efficient sound field capture, which limits their flexibility in practical scenarios. We propose a deep learning (DL)-based approach, leveraging a two-stage network architecture for encoding circular microphone array signals into second-order Ambisonics (SOA) in multi-speaker environments. In addition, we introduce: (i) a novel loss function based on spatial power maps to regularize inter-channel correlations of the Ambisonic signals, and (ii) a channel permutation technique to resolve the ambiguity of encoding vertical information using a horizontal circular array. Evaluation on simulated speech and noise datasets shows that our approach consistently outperforms traditional signal processing (SP) and DL-based methods, providing significantly better timbral and spatial quality and higher source localization accuracy. Binaural audio demos with visualizations are available at \url{https://bridgoon97.github.io/NeuralAmbisonicEncoding/}.

\end{abstract}


\section{Introduction}
\label{sec:intro}

Ambisonics \cite{gerzon1973periphony} is a widely used spatial audio format for capturing, synthesizing, and rendering sound fields. It leverages spherical harmonic (SH) \cite{rafaely2015fundamentals} as basis functions to decompose a sound field into Ambisonic channels of different orders, each of which encodes distinctive spatial information. In general, sound fields captured by spherical microphone arrays are encoded as Ambisonic signals through linear transformations. These encoded signals can be used to reproduce the sound field accurately, either through loudspeakers arranged in specific configurations or can be rendered for headphones via binaural processing. To ensure high-fidelity spatial audio reproduction, it is essential to minimize errors introduced during the Ambisonic encoding process \cite{moreau20063d}.

The effectiveness of Ambisonic encoding is fundamentally limited by the microphone array used in practice. Ideally, numerous microphones uniformly distributed on a sphere are required to obtain Ambisonic signals with high spatial resolution. However, practical microphone arrays often have limited capsules distributed irregularly, depending on the application scenario (e.g., wearable devices \cite{mccormack2022parametric}). This can lead to issues such as spatial aliasing \cite{lubeck2023spatial} and poor spatial coverage of the captured sound field, ultimately degrading the sound field fidelity after encoding. To mitigate the adverse effects of practical microphone arrays, existing studies have explored different Ambisonic encoding approaches. These include traditional least squares (LS)-based optimization using the steering vectors of microphone arrays \cite{gao2018high,bastine2022ambisonics,gayer2024ambisonics}, adding constraints to regularize the orthogonality of Ambisonic channels \cite{schorkhuber2017ambisonic}, and parameterizing the sound field in terms of source directions and diffuseness \cite{mccormack2022parametric}. 

Recently, deep learning (DL) has been applied to Ambisonic encoding \cite{gao2022sparse,heikkinen2024neural} and other tasks related to spatial audio, such as generating spatial audio from mono microphone recordings \cite{morgado2018self,rana2019towards,lim2024enhancing}, binaural rendering from Ambisonics \cite{zhu2024end}, upsampling Ambisonics to higher orders \cite{routray2019deep}, and estimating virtual microphone signals from existing arrays \cite{ochiai2021neural}. The spatial audio generation methods leverage visual data to learn the spatial distribution of sound sources, facilitating the Ambisonic encoding process. However, encoding from only microphone signals is challenging, as it requires implicit extraction of spatial cues from the amplitude and phase information captured by the microphone array(s).

In \cite{gao2022sparse,heikkinen2024neural}, convolution-based DNNs are adopted to learn the transformation from the microphone signals to Ambisonic signals. In \cite{gao2022sparse}, the DNN consists of convolutional layers for different frequency bands, with an additional $l_1$-norm penalty in the loss function to enforce network sparsity. In \cite{heikkinen2024neural}, the DNN is adapted from the U-net architecture \cite{ronneberger2015u}, with channel-wise coherence and energy-based regularization introduced into the loss function. Although these methods have shown similar or better encoding performance compared to traditional LS-based methods under regular microphone array geometries, they may not generalize well to other array layouts in practice, as the network architectures and loss functions used are not tailored for the encoding problem as well as the intrinsic properties of Ambisonic signals (e.g., orthogonality). 

\begin{figure}[tbp]
    \centering
    \includegraphics[width=\linewidth]{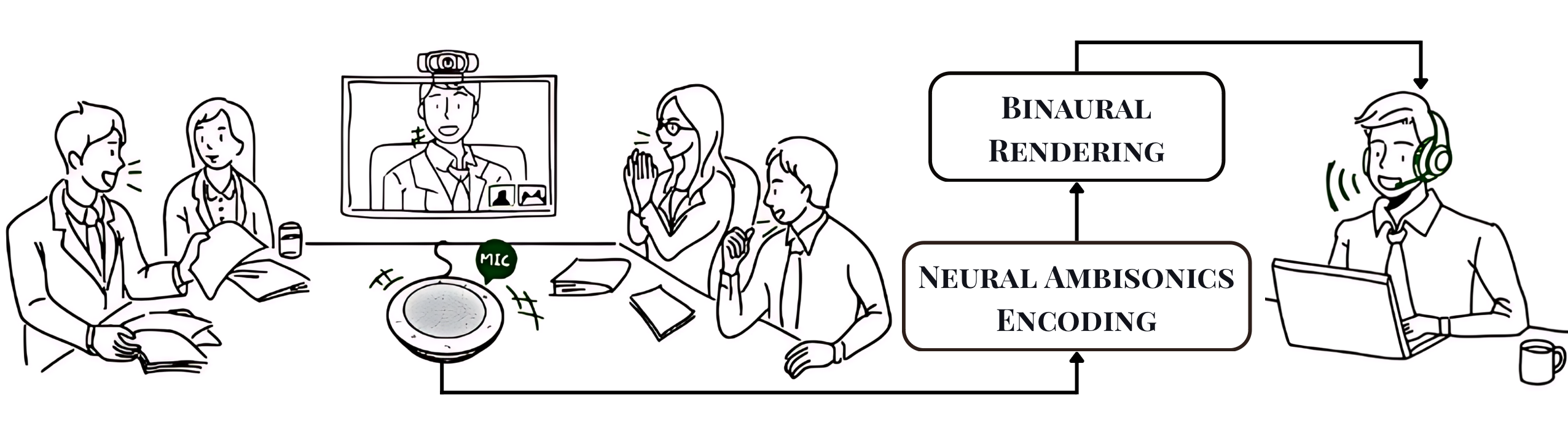}
    \vspace{-2em}
    \caption{Illustration of using Ambisonic encoding for teleconferencing.}
    \label{fig:application}
    \vspace{-2em}
\end{figure}

In this paper, we aim to further improve the performance of DL-based Ambisonic encoding under more challenging conditions, for multi-speaker scenarios such as teleconferencing (Fig.~\ref{fig:application}). Specifically, we choose a circular microphone array situated on the horizontal plane to encode full-3D Ambisonic signals. To guide the DNN in learning the spatial structures of the sound field, we propose a two-stage network architecture that mimics the processes of plane wave decomposition and Ambisonics synthesis. Additionally, we introduce a spatial power map-based loss function to regularize the inter-channel correlation of Ambisonic signals. To address the ambiguity of encoding vertical sound field information using horizontal microphone arrays, we introduce a channel permutation process that discriminates the upper and lower half-space at model inference. We evaluate our proposed method against existing SP- and DL-based encoding methods for: (i) timbral audio quality, (ii) spatial audio quality, and (iii) source localization accuracy.

\section{Problem Formulation}
\label{sec:problem}

In the general problem of Ambisonic encoding, we consider a model consisting of $M$ omnidirectional microphones placed in the free field at $(r_m, \theta_m, \phi_m)$, $m=1,\cdots,M$, and plane waves from $Q$ directions $(\theta_q, \phi_q)$, $q=1,\cdots,Q$. The signals received by the microphones at frequency $f$, $\mathbf{x}(f)$, are related to the signals (or source strengths) associated with the plane waves, $\mathbf{s}(f)$, and the microphone noise signals, $\mathbf{n}(f)$, as
\begin{equation}
    \mathbf{x}(f) = \mathbf{V}(f)\mathbf{s}(f)+\mathbf{n}(f),
\end{equation}
where $\mathbf{x}(f) = [x_1(f),\cdots,x_M(f)]^T$, $\mathbf{s}(f)=[s_1(f),\cdots, \\ s_Q(f)]^T$, $\mathbf{n}(f) = [n_1(f),\cdots,n_M(f)]^T$, and $\mathbf{V}(f)\in \mathbb{C}^{M\times Q}$ denotes the array steering matrix, with each element $V_{mq}$ referring to the transfer function from the $q$-th plane wave to the $m$-th microphone. These plane waves are represented in the Ambisonic domain by utilizing the spherical harmonics (SH) functions as
\begin{equation}
    \mathbf{b}_N(f) = \mathbf{Y}^H_N \mathbf{s}(f),
\end{equation}
where $\mathbf{b}_N(f)=[b_{00}(f),\cdots,b_{NN}(f)]^T$ are the encoded Ambisonic signals of order $N$, $\mathbf{Y}_N = [\mathbf{y}_{00},\cdots,\mathbf{y}_{NN}]\in \mathbb{C}^{Q\times (N{+}1)^2}$ is SH matrix for the given plane waves, and $\mathbf{y}_{nm} = [Y_{nm}(\theta_1,\phi_1),\cdots,Y_{nm}(\theta_Q,\phi_Q)]^T$, where $Y_{nm}(\theta_q,\phi_q)$ is the SH function of order $n$ and degree $m$ corresponding to the angle $(\theta_q,\phi_q)$. 

This study aims to develop a DL-based end-to-end system ($\mathcal{F}$) that transforms the microphone signals $\mathbf{x}(f)$ to an approximation of the ideal Ambisonic signals $\hat{\mathbf{b}}(f)$ for arbitrary source directions:
\begin{equation}
    \mathcal{F}: \mathbf{x}(f) \mapsto \mathbf{b}(f).
    \label{eq:mapping}
\end{equation}

\section{Proposed Neural Ambisonic Encoding DNN}

\begin{figure*}[tbp]
    \centering
    \includegraphics[width=0.9\textwidth]{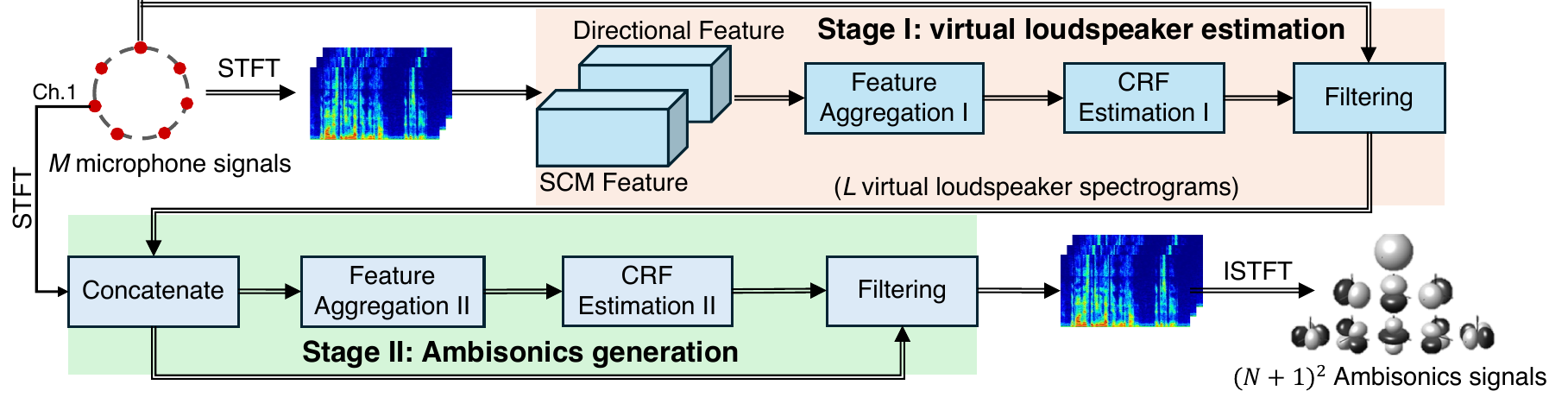}
    \caption{Proposed network architecture for Ambisonic encoding.}
    \label{fig:architecture}
\end{figure*}

The overall architecture of the proposed Ambisonic encoder DNN is depicted in Fig.~\ref{fig:architecture}, which consists of two stages jointly trained in an end-to-end approach: (i) virtual loudspeaker signal estimation and (ii) Ambisonics generation. The DNN input is the stacked $M$-channel microphone signals. After the short-time Fourier transform (STFT), the time-frequency domain signals, $\mathbf{X}(t,f)$, are used to extract two types of audio features: the directional feature \cite{gu20213d}, which is the cosine difference between the inter-channel phase difference and target-dependent phase difference, and the spatial covariance matrix (SCM) of the microphone signals. 

\subsection{Stage I: Virtual Loudspeaker Signal Estimation}
In this stage, the DNN uses the extracted features to predict complex-valued ratio filters (cRFs) for estimating the virtual loudspeaker signals corresponding to sound sources captured by the microphone array. This process is conceptually similar to plane wave decomposition of the captured sound field \cite{rafaely2004plane}. The extracted features are first processed through a one-layer long short-term memory (LSTM) module, a multi-head self-attention (MHSA) \cite{NIPS2017_3f5ee243} module, and a linear layer for feature aggregation. The aggregated features are then passed through a series of four 1-D time-domain convolutional layers (T-Conv) to estimate the cRF masks, $\mathrm{cRF}_\text{LS}\in \mathbb{C}^{L\times M}$, for generating $L$ virtual loudspeaker signals, $\mathrm{VLS}(t,f) \in \mathbb{C}^L$:
\begin{equation}
\mathrm{VLS}(t,f)=\sum_{i\in[-I,I]}\mathrm{cRF}_{\text{LS}}(t,f,i) \mathbf{X}(t{+}i,f),
\end{equation}
where $i$ represents the taps of the cRFs. 

\subsection{Stage II: Ambisonics Generation}

In this stage, the DNN uses the estimated virtual loudspeaker signals and a concatenated single-channel microphone signal (see Fig.~\ref{fig:architecture}) to estimate a nonlinear spatial transformation for generating Ambisonic signals. Incorporating the microphone signal helps the DNN accurately generate the zeroth-order (omnidirectional) Ambisonic signal, similar to how residual connections in deep learning preserve key information and stabilize training. The concatenated signals are passed through linear layers for dimension reduction and gated recurrent unit (GRU) layers for feature aggregation, followed by four narrow-band blocks from SpatialNet \cite{quan2024spatialnet} to estimate another set of cRFs. Each block includes a MHSA module and a time-convolutional feedforward module (T-ConvFFN), which perform spatial clustering and temporal smoothing/filtering, respectively. The MHSA also helps adapt the DNN weights to each loudspeaker’s spatial position. For more details on the narrow-band block, see \cite{quan2024spatialnet}. The Ambisonic signals, $\hat{\mathbf{B}}(t,f)$, are generated by filtering the concatentaed signals with the estimated cRFs, $\mathrm{cRF}_\text{SH}\in\mathbb{C}^{(N+1)^2\times(L+1)}$:
\begin{align}
\resizebox{0.9\linewidth}{!}{$
\begin{aligned}
    \hat{\mathbf{B}}(t,f)=\sum_{i \in [-I',I']}\mathrm{cRF_{SH}}(t,f,i)[\mathrm{VLS}(t{+}i,f)^T, X_1(t{+}i,f)]^T.
\end{aligned}
$}
\end{align}
Next, energy normalization is employed on the estimated Ambisonic signals to ensure the energy of the estimated zeroth-order Ambisonics matches that of the microphone signal. Finally, $\hat{\mathbf{B}}(t,f)$ is transformed to the time domain, $\hat{\mathtt{b}}$, with the inverse-STFT operation. 

\subsection{Loss functions}\label{ssec:losses}

We use four loss functions to train the proposed DNN. The first three, Magnitude $l_1$-norm loss, signal-invariant signal-to-noise ratio (SI-SNR, \cite{le2019sdr}) loss, and coherence loss \cite{heikkinen2024neural}, aim to minimize the channel-wise errors between the estimated and ground truth Ambisonics. They are defined as
\vspace{-1mm}
\begin{align}
\resizebox{0.8\linewidth}{!}{$
\begin{aligned}
    \mathcal{L}_\text{MagMAE} =\frac{1}{(N{+}1)^2 TF} \sum_{n=1}^{(N{+}1)^2} \sum_{t=1}^T \sum_{f{=}1}^F ||B_n(t,f)|{-}|\hat{B}_n(t,f)||,
\end{aligned}
$}
\end{align}
\vspace{-1mm}
\begin{align}
\resizebox{0.8\linewidth}{!}{$
\begin{aligned}
    \mathcal{L}_\text{SI-SNR} &= - \frac{1}{(N{+}1)^2} \sum_{n=1}^{(N{+}1)^2} 10 \log_{10} \frac{\|e_{\text{target},n}\|^2}{\|e_{\text{res},n}\|^2} \\  &= -\frac{1}{(N{+}1)^2} \sum_{n=1}^{(N{+}1)^2} 10 \log_{10}\frac{\|\frac{\langle\hat{\mathtt{b}}_n, \mathtt{b}_n\rangle}{\|\mathtt{b}_n\|^2}\mathtt{b}_n \|^2}{\|\frac{\langle\hat{\mathtt{b}}_n, \mathtt{b}_n\rangle}{\|\mathtt{b}_n\|^2}\mathtt{b}_n-\hat{\mathtt{b}}_n\|^2},
\label{eq:loss_sisnr}
\end{aligned}
$}
\end{align}
where $\hat{\mathtt{b}}_n$ and $\mathtt{b}_n$ are the estimated and ground truth time-domain Ambisonic signal of channel $n$, respectively, and
\begin{align}
\resizebox{0.9\linewidth}{!}{$
\begin{aligned}
    \mathcal{L}_\text{Coherence} &= 1 - \sum_{n=1}^{(N{+}1)^2}\sum_{f=1}^F \frac{|\sum_{t=1}^T b_n^{*}(t,f) \hat{b}_n(t,f)|^2}{((N{+}1)^2F) \sum_{t'=1}^T |b_n(t',f)|^2 \cdot \sum_{t''=1}^T |\hat{b}_n(t'',f)|^2}.
\end{aligned}
$}
\label{eq:loss_coherence}
\end{align}

The fourth loss function is based on the spatial power map \cite{mccormack2017parametric} derived from Ambisonic signals. The power map, $\Gamma(\theta,\phi)$, is computed using fixed beamformer weights, $\mathbf{w}(\theta,\phi)\in \mathbb{C}^{(N{+}1)^2}$, corresponding to directions, $\{\theta_i,\phi_i\}_{i=1}^{Q}$:
\begin{equation}
     \Gamma(\theta_i,\phi_i) = \|\mathbf{w}^H(\theta_i,\phi_i)\mathtt{b}\|.
\end{equation}
In this work, we choose $\mathbf{w}(\theta,\phi)$ to be the maximum directivity index (max-DI) beamformer weights \cite{lluis2023direction} for 1296 directions (spherical design of degree 50 \cite{graf2011computation}). The power map loss minimizes the difference between the estimated power map, $\hat{\Gamma}(\theta,\phi)$, and the ground truth power map, $\Gamma(\theta,\phi)$:
\begin{equation}
    \mathcal{L}_{\text{PowerMap}} = \alpha \mathcal{L}_{\text{KL}}(\Gamma,\hat{\Gamma}) + \beta \frac{1}{Q}\sum_{i=1}^{Q}|\Gamma(\theta_i,\phi_i) - \hat{\Gamma}(\theta_i,\phi_i)|^2,
\end{equation}
where $\mathcal{L}_{\text{KL}}$ is the Kullaback-Leibler divergence between the power map distributions, and $\alpha,\beta$ are weighting parameters ($\alpha{=}1,\beta{=}100$ used in training). The loss definition is inspired from \cite{wu2021binaural} where audio localization maps are considered. The power map loss helps regularize the cross-channel correlation and therefore enhances the prediction of spatial information. During training, these four loss functions are equally weighted and combined.

\subsection{Vertical channel permutation}\label{ssec:permutation}
When using a horizontal microphone array to encode Ambisonics, the vertical sound field components cannot be accurately encoded due to the lack of ability to distinguish between the sound sources ``mirrored'' in the horizontal plane. However, if we assume that all the sound sources are located in the same half-space, and this positioning is known or can be assumed, such ambiguity can be addressed by permuting the vertical Ambisonic channels. This assumption is often valid in scenarios like teleconferencing, where most speakers are either above or below the microphone array. Specifically, we permute the vertical channels (e.g., for second-order Ambisonics (SOA), $y_{1,0}, y_{2,-1},y_{2,0},y_{2,1}$) by multiplying them with -1 (i.e., inverting the phase, equivalent to vertically flipping the sound sources) at model inference when the sources are in the lower half-space. This ensures a one-to-one mapping from the microphone input to the estimated sound field.

\section{Experimental Setup}

We use an open microphone array with eight omnidirectional microphone capsules arranged in a circle with a 5-cm radius on the horizontal plane for capturing the sound field. The target Ambisonics is SOA (9 channels). The STFT/iSTFT used in the model has an FFT size of 512 and a hop size of 256, assuming a 16 kHz sampling frequency. The directional feature is computed with 128 uniformly sampled directions and 6 microphone pairs, and the SCM feature contains 128 channels (real and imaginary parts of the 64-channel SCM concatenated). In the first stage, the MHSA module has 8 heads, and the T-Conv modules use kernels with size 3, stride 1, dilation 1, and padding 1 on both sides. The number of virtual loudspeaker channels is set to 50. In the second stage, the narrow-band block parameters follow the ``SpatialNet-small'' preset from \cite{quan2024spatialnet}. The cRFs in both stages are set to $I{=}I'{=}2$ for the filter taps. The channel dimension of all the intermediate layers is 256, resulting in a total model size of 8.1 M.

We simulate an audio dataset with speech scenarios involving 1, 2, and 3 speakers for training and evaluation. The speakers are 1 m from the array center, with directions randomly sampled on a sphere; the elevation range is limited to $[30^{\circ},150^{\circ}]$ as speakers outside this range are uncommon in practical applications. In multi-speaker scenarios, speakers are in the same half-space, and the upper/lower information is known for the permutation operation. Speech signals are simulated by convolving dry signals with room impulse responses (RIRs) for each microphone using the FRAM-RIR method \cite{luo2024fast}. The room size is randomized between $[4,4,3.5]m$ and $[10,10,6]m$, and the RT60 is randomly sampled in $[0.05,0.9]s$. The ground truth Ambisonic channels are treated as co-located microphones at the array center, with directivity patterns equivalent to the SH functions ($\mathbf{y}_{nm}$). For each speaker, signals corresponding to the direct sound and secondary reflections are filtered individually based on the source locations and summed together. The speech dataset includes 158 hours for training, 8 hours for validation, and 2 hours for evaluation. A noise dataset that contains single sources evenly distributed in the space with white noise is also simulated to evaluate source localization accuracy in the estimated Ambisonic signals.

\section{Evaluation}

\begin{table*}[t]
\centering
\resizebox{\textwidth}{!}{
\begin{tabular}{lccccccccc}
\toprule
\multirow{4}{*}{\textbf{Methods}} & \multicolumn{7}{c}{\textbf{Audio Quality}} & \multicolumn{2}{c}{\textbf{Source Localization Accuracy}} \\ \cmidrule(lr){2-8} \cmidrule(lr){9-10}
 & \multicolumn{4}{c}{\textbf{Timbral}} & \multicolumn{3}{c}{\textbf{Spatial}} & \multirow{3}{*}{\textbf{ERR\_AZI ($^{\circ}$)$\downarrow$}} & \multirow{3}{*}{\textbf{ERR\_ELEV ($^{\circ}$)$\downarrow$}} \\ \cmidrule(lr){2-5} \cmidrule(lr){6-8}
 & \multicolumn{2}{c}{\textbf{Temporal}} & \multicolumn{2}{c}{\textbf{Spectral}} & \textbf{SH} & \multicolumn{2}{c}{\textbf{Binaural}}  &  \\ \cmidrule(lr){2-3} \cmidrule(lr){4-5} \cmidrule(lr){6-6} \cmidrule(lr){7-8}
 & \textbf{SI-SNR (dB)$\uparrow$} & \textbf{ENV$\downarrow$} & \textbf{LSD (dB)$\downarrow$} & \textbf{Coherence$\uparrow$} & \textbf{RMSE\_MAP$\downarrow$} & \textbf{RMSE\_ILD (dB)$\downarrow$} & \textbf{RMSE\_IC$\downarrow$} &  &  \\ 
\midrule
DSB + Ambi. Synth. & -2.842 & 0.077 & 1.209 & 0.376 & 0.270 & 2.152 & 0.563 & 6.651 & 52.601 \\
LS-based Filtering I & 2.677 & 0.063 & 0.776 & 0.434 & 0.169 & 1.908 & 0.585 & 2.857 & 40.307 \\
LS-based Filtering II & 3.141 & 0.063 & 0.776 & 0.434 & 0.141 & 1.904 & 0.585 & 2.866 & 12.259 \\
U-net-based (8.4M) \cite{heikkinen2024neural} & 3.283 & 0.060 & 0.776 & 0.534 & 0.139 & 1.738 & 0.258 & 20.847 & 30.162 \\
SpatialNet (8.3M) \cite{quan2024spatialnet} & 6.351 & 0.051 & 0.503 & \textbf{0.626} & 0.122 & 1.302 & 0.316 & 16.845 & 32.361 \\
\hline\hline
\textbf{Proposed DL-system} (8.1M) & \textbf{6.353} & \textbf{0.042} & \textbf{0.350} & 0.610 & \textbf{0.072} & \textbf{1.043} & \textbf{0.140} & \textbf{1.555} & \textbf{10.204} \\
\hline\hline
(w/o PowerMap loss) & \textit{6.781} & \textit{0.045} & \textit{0.470} & \textit{0.620} & \textit{0.118} & \textit{1.469} & \textit{0.231} & \textit{2.515} & \textit{13.814} \\
(w/o Perm. \& PowerMap loss) & \textit{6.345} & \textit{0.040} & \textit{0.314} & \textit{0.616} & \textit{0.138} & \textit{1.200} & \textit{0.132} & \textit{18.752} & \textit{32.215} \\
\bottomrule
\end{tabular}}
\vspace{0.1em}
\caption{Comparison of Ambisonic encoding methods across metrics regarding audio quality and source localization accuracy.}
\label{tab:model_comparison}
\end{table*}
\begin{figure*}[tbp]
    \centering
    \includegraphics[width=\textwidth]{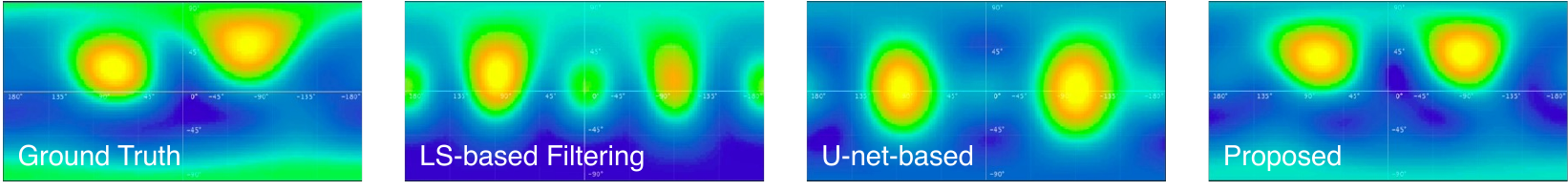}
    \caption{Spatial power map (plotted using the SPARTA plugin \cite{mccormack2019sparta}) for ground truth and estimated Ambisonics using different encoding approaches.}
    \label{fig:powermap}
\end{figure*}

\subsection{Metrics}

We evaluate two aspects of Ambisonic encoding performance using the speech and noise datasets: audio quality (timbral and spatial) and source localization accuracy. For timbral audio quality, we choose SI-SNR (Eq.~\ref{eq:loss_sisnr}) and envelope distance (ENV, \cite{morgado2018self}) between the estimated ($\hat{\mathtt{b}}$) and ground truth ($\mathtt{b}$) Ambisonic signals after Hilbert transform to measure temporal quality. Log spectral distance (LSD, \cite{gray1976distance}) and channel-wise coherence (Eq.~\ref{eq:loss_coherence}) between $\hat{\mathbf{b}}(f)$ and $\mathbf{b}(f)$ are used for spectral quality. For spatial quality, we calculate the root mean square error (RMSE) between the power map $\hat \Gamma(\theta,\phi)$ and $\Gamma(\theta,\phi)$ in the SH domain; in the binaural domain, following \cite{mccormack2022parametric}, we decode Ambisonics into binaural signals using head-related transfer functions (HRTFs) from the KEMAR dataset \cite{gardner1995hrtf}), then compute RMSE for inter-aural level difference (ILD) and inter-aural coherence (IC). For source localization accuracy, using the noise dataset, we estimate the direction of arrival by finding the maximum RMS value of the spatial power map, $\hat \Gamma(\theta,\phi)$, from the estimated Ambisonics and then calculate the azimuth and elevation errors relative to the ground truth. For simplicity, all metrics are computed for SOA and averaged across time and frequency for all evaluation samples.

\subsection{Baseline methods}

We implement four baseline encoding methods (two SP-based and two DL-based) for performance comparison with the proposed method. The first SP-based method uses a delay-and-sum beamformer (DSB) to generate virtual loudspeaker signals for 160 fixed directions and synthesizes Ambisonics by filtering each loudspeaker signal with corresponding SH functions. The second SP-based method, similar to the conventional baselines in \cite{gao2018high,mccormack2022parametric,heikkinen2024neural}, filters the microphone signals using an encoding matrix, $\mathbf{E} \in \mathbb{C}^{(N{+}1)^2\times M}$, which linearly transforms the signals as
\begin{equation}
    \hat{\mathbf{b}}(f) = \mathbf{E}(f) \mathbf{x}(f),
\end{equation}
where $\hat{\mathbf{b}}(f)$ is the estimated Ambisonic signals. A closed-form solution for $\mathbf{E}(f)$ can be obtained in a least-squares sense, by minimizing the expectation of the $l_2$-norm difference between $\hat{\mathbf{b}}(f)$ and $\mathbf{b}(f)$\cite{moreau20063d,politis2017comparing}:
\begin{equation}
    \mathbf{E}(f) = \mathbf{Y}_N^H(f)\mathbf{V}^H(f)[\mathbf{V}(f) \mathbf{V}^H(f) + \lambda(f)\mathbf{I}_M]^{-1},
\label{eq:enc_matrix}
\end{equation}
where $\lambda(f)$ is a regularization parameter set to 0.01 in our experiment. The steering matrix $\mathbf{V}$ is sampled at 5$^{\circ}$ resolution. To address the ambiguity issue (Sec.~\ref{ssec:permutation}), when it is unknown whether the sound sources are above or below the horizontal plane, the encoding matrix is generated using only the steering vectors above the horizontal plane. When such information is known, separate encoding matrices are generated for both cases using the corresponding steering vectors. The third method adopts the U-net-based architecture \cite{heikkinen2024neural} with unchanged hyperparameters and loss functions, while the fourth method adopts the ``SpatialNet-big'' preset from SpatialNet \cite{quan2024spatialnet}, employing all the loss functions from Sec.~\ref{ssec:losses} except the spatial power map loss. No channel permutation is applied to these DNN models.

\subsection{Results}

Table \ref{tab:model_comparison} shows the performance of the four baseline methods and the proposed method (including two in ablation settings) in terms of audio quality and source localization accuracy. The second and third rows refer to single LS-based filtering (I) using steering vectors above the horizontal plane, and two filters (II) generated separately for above and below, respectively. 

Regarding audio quality, the DSB-based method performs worst in both timbral and spatial quality, likely due to the frequency-dependent energy roll-off (which leads to signal coloration) and low spatial resolution outside the horizontal plane. LS-based filtering improves audio quality, especially in phase-related metrics such as SI-SNR, with further gains when separate filters are used. The DL-based baselines outperform SP-based ones as discretization artifacts are avoided, and SpatialNet outperforms U-net, possibly due to its narrow-band processing. Our proposed model performs similarly or better in timbral quality and SH-domain spatial quality, with significantly better binaural spatial quality, likely due to the addition of the power map loss. For source localization accuracy, SP-based baselines, despite having lower audio quality, achieve better azimuth prediction than DL-based ones. This indicates that DNNs trained with only channel-wise loss functions may not fully preserve spatial information. In addition, using separate LS-based filters reduces the elevation error, confirming the ambiguity between upper and lower half-spaces. In comparison, our proposed method yields the lowest localization errors for both azimuth and elevation angles. The impact of adding channel permutation and the power map loss is analyzed by comparing the last three rows, where we see significant improvements in spatial quality and localization accuracy, with slight degradation in timbral quality. Compared to SpatialNet, our model without permutation and the power map loss still yields better ENV, LSD, and binaural quality, with similar performance in other metrics.

Fig.~\ref{fig:powermap} shows the spatial power maps computed from the ground truth and estimated Ambisonics, for a segment of a two-speaker scenario. The proposed method demonstrates better preservation of the spatial information in the captured sound field, while the two baseline methods introduce artifacts, such as “ghost” source images and shifts in the source positions.

\section{Conclusion and future work}
In this paper, we presented a DL-based Ambisonic encoding method designed for multi-speaker scenarios, such as teleconferencing, using a circular microphone array. The two-stage network architecture mimics plane wave decomposition and Ambisonics synthesis, incorporating channel permutation and a spatial loss function to enhance spatial information preservation. Evaluation with simulated speech and noise datasets demonstrated that our method significantly improves spatial audio quality and source localization accuracy compared to existing baseline methods. Future work could explore the impact of different microphone array layouts and further optimize the network architecture.

\newpage
\bibliographystyle{IEEEtran}
\bibliography{IEEEabrv,refs}


\end{document}